\documentclass[12pt]{article}
\usepackage{amsmath,amssymb,amsfonts}
\begin{document}

\begin{titlepage}
                            \begin{center}
                            \vspace*{-1cm}
        \large\bf The Dirac monopole and differential characters.\\

                            \vfill

              \small\sf NIKOS \  KALOGEROPOULOS\\

                            \vspace{0.2cm}

 \small\sf Department of Science,\\
               BMCC  --  The City University of New York,\\
              199 Chambers St., \  New York, NY 10007, USA\\
                            \end{center}

                            \vfill

                     \centerline{\normalsize\bf Abstract}
                            \vspace{0.3cm}
\setlength{\baselineskip}{24pt}
 \normalsize\rm\noindent We describe
the Dirac monopole using the Cheeger-Simons differential
characters. We comment on the r\^{o}le of the Dirac string and on
the connection with Deligne cohomology.

                             \vfill


                             \vfill

\noindent\rule{8cm}{0.2mm}\\
\begin{tabular}{ll}
\small\rm E-mail: & \small\rm nkalogeropoulos@bmcc.cuny.edu\\
                  & \small\rm nkaloger@yahoo.com\\
\end{tabular}
\end{titlepage}


                                 \newpage

\setlength{\baselineskip}{24pt}

\noindent The apparent lack of symmetry of the Maxwell equations
$\nabla\!\cdot\!{\bf E}\!=\!\rho_E$ and $\nabla\!\cdot{\bf
B}\!=\!0$ under the duality transformation ${\bf
E}\!\rightarrow\!{\bf B}, \ {\bf B}\!\rightarrow\! -{\bf E}$
inspired Dirac [1], [2] to study the consequences of the existence
of a hypothetical magnetic charge density $\rho_M$. For
$\rho_M\!=\!4\pi g \delta (r)$, by analogy to Coulomb's law, the
magnetic field produced would be ${\bf B}\!=\!gr^{-3}{\bf r}$. The
magnetic flux on a 2-sphere $S^2$ surrounding that magnetic charge
would then be
\begin{equation}
\Phi\!=\!\int_{S^2}\!{\bf B}\!\cdot\!{\bf dS}\!=\!4\pi g
\end{equation}
Classical electrodynamics can also be expressed in terms
of the vector potential ${\bf A}$ defined by ${\bf B}\!=
\!\nabla\!\times\!{\bf A}$. Then
\begin{equation}
\Phi\!=\!\int_{S^2}\!{\bf B}\!\cdot\!{\bf dS}\!=\!
          \int_{\partial S^2}\!{\bf A}\!\cdot\!{\bf dl}\!=\!0
\end{equation}
if ${\bf A}$ is assumed smooth everywhere on $S^2$. The apparent
contradiction of (1) and (2) compels somebody to accept that
either there is no globally defined choice for ${\bf A}$ or that
the vector potential ${\bf A}$ must have a line singularity which
starts from the location of the monopole and stretches to
infinity. Most of the treatments of this system take the former
view since the latter introduces a line singularity (``Dirac
string'') which is non-physical, as can be demonstrated by the
fact that a gauge transformation changes its location. We present
an alternative  description of the Dirac string  by using the
Cheeger-Simons differential characters. These objects incorporate
singularities so they generalize singular cohomology, thus
providing more flexible algebraic objects useful in the
description of topological defects. \\

\noindent The conventional approach  is to model the Dirac string
as a principal $U(1)$  bundle over $S^2$. The magnetic monopole
charge $g$ distinguishes the different ways this vector bundle can
be twisted, so it represents an element of
$H^2(S^2,\!\mathbb{Z})$. The generator of this group, when the
corresponding $1$st Chern number of the bundle is $\pm 1$, is the
first integral Chern class,  which provides a topological
obstruction to the  bundle being trivial. The vector potential
${\bf A}$ induces an isomorphism
$H^2(S^2,\!\mathbb{Z})\!\rightarrow\!H^2(S^2,\!\mathbb{R})$. The
generator of the latter group, when the corresponding $1$st Chern
number of the bundle is equal to $\pm 1$, is the first real Chern
class. This provides an obstruction to the corresponding potential
being globally flat, namely $dA=0$ for the $u(1)$-valued 1-form
$A$ corresponding to ${\bf A}$. Since the two cohomology groups
are isomorphic, due to the torsion-free nature of
$H^2(S^2,\!\mathbb{Z})$ the two obstructions are
equivalent and are used interchangeably [3].\\

 Instead of taking the conventional approach, we may insist on
the existence of a globally defined vector potential ${\bf A}$.
Then ${\bf A}$, as well as the corresponding connection $A$, will
have to have line singularities. The theory of singular
connections has been studied extensively in [4]. Consider a
singular $u(1)$-valued one-form $A$ which is defined on the
complement of a point in $\mathbb{R}^3$ i.e. on
$\mathbb{R}^3\!-\!\{0\}$, the monopole location being excluded
from $\mathbb{R}^3$ and let $F=dA$ be the corresponding curvature.
Consider also two cycles $C$ and $C^{\prime}$ on
$\mathbb{R}^3\!-\!\{0\}$ that do not pass through the monopole.
Let $S$ be a 2-chain such that $C^{\prime}\!=\!C\!+\!\partial S$.
Here $C$ and $C^{\prime}$ are simplicial models of the circle, $S$
is a simplicial model of a 2-dimensional surface and $\partial$
denotes the simplicial boundary operator. The simplicial model of
the surface of interest should be a 2-chain, and not a 2-cycle,
otherwise $\partial S=\!\emptyset$. Then
\begin{equation}
\int_{C^{\prime}}\!A\!=\!\int_C\!A+\int_{\partial S}\!F
\end{equation}
When an external electromagnetic field is minimally coupled to
matter, the wave-function describing the matter obtains an extra
phase \ $\exp(i\!\int_S\!A)$. The quantity of physical
significance therefore, is the $\bmod \ \mathbb{Z}$ reduction of
(3). We are interested, therefore, in singular connections that
obey the relation
\begin{equation}
\int_{C^{\prime}}\!A\!=\!\int_C\!A+\int_{\partial S}\!F \;\; \bmod\mathbb{Z}
\end{equation}
This, by definition, means that $A$ is an element of the first
differential character group
$\stackrel{\wedge\;}{H^1}\!(\mathbb{R}^3-\{0\},\!\mathbb{R}/\!\mathbb{Z})$.
A differential character then, is an object that can be defined
(non-canonically) by a differential form with singularities [5].
The non-canonical realization encodes the non-uniqueness, and the
subsequent lack of physical meaning, of the Dirac string. This
construction can be straightforwardly generalized to obtain
k-dimensional differential characters. These  are elements of the
k-th differential character group
$\stackrel{\wedge\;}{H^k}\!(\mathbb{R}^3\!-\!\{0\},\mathbb{R}/\mathbb{Z})$
[6]. Since $S^2$ is a deformation retract of
$\mathbb{R}^3\!-\!\{0\}$ [7], we can use $S^2$ instead of
$\mathbb{R}^3\!-\!\{0\}$ in the calculation of the cohomology
groups. It turns out that there is an exact sequence [6]
\begin{equation}
0\longrightarrow H^k(S^2,\mathbb{R}/\mathbb{Z})\longrightarrow
     \stackrel{\wedge }{H^k}(S^2,\mathbb{R}/\mathbb{Z})
 \longrightarrow \Lambda_{o}^{k+1}(S^2)\longrightarrow 0
\end{equation}
where $\Lambda_{o}^{k}(S^2)$ represents the closed $k$-forms on
$S^2$ with integral periods. On $S^2$ all 3-forms are trivial.
Therefore, for $k\!=\!2$ the exact sequence (5) implies the
isomorphism
\begin{displaymath}
H^2(S^2,\mathbb{R}/\mathbb{Z})\cong
                     \stackrel{\wedge\;}{H^2}(S^2,\mathbb{R}/\mathbb{Z})
\end{displaymath}

\noindent This is a generic characteristic the highest degree
differential character groups, namely they reduce to the
corresponding singular cohomology group of the same degree. The
universal coefficient theorem gives
\begin{displaymath}
 H^2(S^2, \mathbb{R}/\mathbb{Z}) = Hom (H_2(S^2,\mathbb{Z}))
                    + Ext (H^1(S^2,\mathbb{Z}), U(1))
\end{displaymath}
which reduces to
\begin{displaymath}
H^2(S^2, \mathbb{R}/\mathbb{Z}) = Hom (\mathbb{Z}, U(1))
\end{displaymath}

\noindent One can also consider [5], [6] the exact sequence
\begin{displaymath}
 0\longrightarrow
\Lambda^1(S^2)/\Lambda_o^1(S^2)\longrightarrow
      \stackrel{\wedge }{H^1}(S^2,\mathbb{R}/\mathbb{Z})\longrightarrow
        H^2(S^2,\mathbb{Z})\longrightarrow 0
\end{displaymath}
where $\Lambda^1, \Lambda_o^1$ are the spaces of closed one-forms
and closed one-forms with integral periods respectively. Since all
closed one-forms on $S^2$ have integral periods, the quotient
space is trivial. This implies the isomorphism
\begin{displaymath}
\stackrel{\wedge }{H^1}(S^2,\mathbb{R}/\mathbb{Z})
                             \cong H^2(S^2,\mathbb{Z})\cong \mathbb{Z}
\end{displaymath}
which reproduces the conventional form of the Dirac quantization condition.\\

 The same ideas can be reexpressed in a slightly different way as
follows [6]: Consider the set
$$R^2(S^2,\mathbb{Z})=\{ (\omega, u)\in \Lambda_o^2\times H^2(M,\mathbb{Z}) |
r(u)=[\omega ] \}$$
with $r$ being the natural map
   $r: H^2(S^2,\mathbb{Z})\rightarrow H^2(S^2,\mathbb{R})$
and $[\omega ]$ is the deRham class of $\omega$. Generically  that
map has a large kernel. In this case the map is injective  due to
lack of torsion of $H^2(S^2,\mathbb{Z})$. Then the short exact
sequence
$$ 0\longrightarrow H^1(S^2,\mathbb{R})/r(H^1(S^2,\mathbb{Z}))\longrightarrow
       \stackrel{\wedge }{H^1}(S^2,\mathbb{R}/\mathbb{Z})\longrightarrow
                     R^2(S^2,\mathbb{Z})
          \longrightarrow 0$$
gives rise to the isomorphism
\begin{equation}
\stackrel{\wedge}{H^1}(S^2,\mathbb{R}/\mathbb{Z})\cong R^2(S^2,\mathbb{Z})
\end{equation}
Equation (6) also shows that the differential characters are
generalizations of the differential forms, since the latter would have
to vanish on the boundary of a simplex.\\

\noindent The geometric interpretation of these facts is the following: It is known
[8] that there is an isomorphism
 $$\stackrel{\wedge }{H^1}(S^2, \mathbb{R}/\mathbb{Z})\cong
                                  H^2(S^2,\mathbb{Z}(2)_D^{\infty})$$
between the Cheeger-Simons differential 1-character group and the
second smooth Deligne
cohomology group with coefficients in the complex of sheaves $\mathbb{Z}(2)_D$.
This complex of sheaves is defined to be the complex
 $$\underline{\mathbb{Z}(2)}_{S^2}\longrightarrow \Lambda^0(S^2)
     \stackrel{d}{\longrightarrow}\Lambda^1(S^2)$$
where $\mathbb{Z}(2)\cong (2\pi i)^2\mathbb{Z}\subset \mathbb{R}$ and
   $\underline{\mathbb{Z}(2)}$ is the
constant sheaf corresponding to $\mathbb{Z}(2)$ [9]. The Deligne
cohomology group $H^2(S^2,\mathbb{Z}(2)_D^{\infty})$ is the set of
isomorphism classes of smooth principal $U(1)$- bundles with
connections over $S^2$ [10]. The curvatures of these connections
are elements of $\Lambda^2_o(S^2)$ since the Bianchi identity
$DF=0$ reduces to $dF=0$ for a connection on a $U(1)$-bundle on
$S^2$, or more simply due to dimensional reasons. Furthermore,
$H^1(S^2, U(1))$ is the set of isomorphism classes of flat
connections on smooth $U(1)$-bundles over $S^2$. The exact
sequence (5), for $k=1$ expresses the fact that any isomorphism
class of $U(1)$-bundles with a connection over $S^2$ can be
obtained as the ``twist'' of a non-trivial $U(1)$-bundle with a
connection, by a flat $U(1)$-connection.\\

 We conclude that the differential characters provide a natural
framework in which to express the Dirac quantization condition and
they elucidate some associated geometric structures. Relatively
recently, a theory of relative Cheeger-Simons differential
characters has  been developed [11]. It would be interesting to
see how this formalism can be used to describe
other topological defects [12].\\

                             \vspace{0cm}

\centerline{\sc References}

                            \vspace{0.1cm}

\noindent
1. P.A.M. Dirac, \  \emph{Proc. Roy. Soc.} {\bf A133}, 60 (1931)\\
2. P. Goddard, D.I. Olive, \ \emph{Rep. Prog. Phys.} {\bf 41}, 1357 (1978)\\
3. C. Soul\'{e}, \  \emph{Asterique} {\bf 819}, 411 (1996)\\
4. F.R. Harvey, H.B. Lawson Jr., \ \emph{Asterisque} {\bf 213}, 1
   (1993), \ \emph{AMS Bull.} {\bf 31}, 54 (1994),
   \ \emph{Amer. Jour. Math.} {\bf 117}, 829 (1995)\\
5. J. Cheeger, \  \emph{Inst. Naz. Al. Math., Symp. Mathem.} {\bf IX}, 441 (1973)\\
6. J. Cheeger, J. Simons, \  \emph{``Differential Characters and
   Geometric Invariants''} in \emph{``Geometry and Topology''}
   J. Alexander, J. Harer (Eds). LNM {\bf 1167}, pp. 50-80, Springer
   1985\\
7. M.J. Greenberg, J.R. Harper, \  \emph{``Algebraic Topology. A First Course''}
   Addison-Wesley (1981)\\
8. H. Esnault, \  \emph{Topology} {\bf 27}, 323 (1988)\\
9. J.L. Brylinski, \ \emph{``Loop spaces, characteristic classes and
   geometric quantization''} Prog. Math. {\bf 107}, Birkhauser 1993\\
10. P. Gajer, \  \emph{Invent. Math.} {\bf 127}, 155 (1997)\\
11. R. Zucchini, \  \emph{``Relative topological integrals and
relative Cheeger-Simons differential characters''}, \  arXiv:hep-th 0010110\\
12. A. Vilenkin, E.P.S. Shellard, \ \emph{``Cosmic Strings and
Other Topological Defects"} Cambridge University Press (1994). \\

                               \vfill
\end{document}